\documentclass[10pt,letterpaper]{article}
\usepackage{opex3}
\usepackage{mathptmx}
\usepackage{amsmath}

\newsavebox{\savepar}

\begin{document}

\title{On the Application of a Monolithic Array for Detecting Intensity-Correlated Photons
Emitted by Different Source Types}


\author{D. L. Boiko}
\address{
Centre Suisse d'Electronique et de Microtechnique SA, 2002, Neuch\^atel, Switzerland}
\email{dmitri.boiko@csem.ch} 

\author{N. J. Gunther}
\address{Performance Dynamics,
4061 East Castro Valley Blvd., Castro Valley, California, USA}
\email{njgunther@perfdynamics.com}

\author{N. Brauer, M. Sergio, C. Niclass}
\address{Ecole Polytechnique F\'ed\'erale de Lausanne, Quantum
Architecture Group, 1015, Lausanne, Switzerland}
\email{nilsbenedict.brauer@epfl.ch, maximilian.sergio@epfl.ch, cristiano.niclass@epfl.ch}

\author{G. B. Beretta}
\address{ HP Laboratories, 1501 Page Mill Road, Palo Alto, California,
USA}
\email{giordano.beretta@hp.com}

\author{E. Charbon}
\address{Technische Universiteit Delft, Mekelweg 4, 2628 CD Delft, The Netherlands}
\email{e.charbon@tudelft.nl}


\begin{abstract*}
It is not widely appreciated that many subtleties are involved in the accurate measurement of
intensity-correlated photons; even for the original experiments of Hanbury Brown and Twiss (HBT).
Using a monolithic 4$\times$4 array of single-photon avalanche diodes (SPADs),
together with an off-chip algorithm for processing streaming data,
we investigate the difficulties of measuring second-order photon
correlations $g^{(2)}(x^{\prime},t^{\prime},x,t)$ in a wide variety of light fields that exhibit dramatically different correlation  statistics: a multimode He-Ne laser, an incoherent intensity-modulated lamp-light source and a thermal light source.
Our off-chip algorithm treats multiple photon-arrivals at pixel-array pairs,
in any observation interval, with photon fluxes limited by detector saturation,
in such a way that a correctly normalized $g^{(2)}$ function is guaranteed.
The impact of detector background correlations between SPAD pixels and afterpulsing effects on second-order coherence measurements is discussed.
These results demonstrate that our monolithic SPAD array enables access to effects that are otherwise impossible to measure with stand-alone detectors.
\newline
\end{abstract*}


\section{Introduction}

The first-order correlation function, $g^{(1)}$, is widely used in
optical applications because a simple experimental arrangement makes it easily accessible,
e.g. Young's double-slit interferometer \cite{Lundeberg07}.
In Young's experiment, two points on the wavefront separated by a distance $x_{12}$
produce a fringe pattern with resultant
intensity $I_1+I_2+2 \sqrt{I_1I_2} | g^{(1)}(\mathbf{x}_{12},\tau)|
\cos(\Delta \varphi_{12})$ at a screen location where the phase difference is $\Delta \varphi_{12}$
 and the propagation time-difference is $\tau$.
 The fringe visibility $\mathcal{V}$ measures the magnitude of the first-order correlation function $| g^{(1)}(\mathbf{x}_{12},\tau)|=\frac{I_1+I_2}{2\sqrt{I_1I_2}} \mathcal{V} $ between field points at the double pinhole. When the light-state distribution, $| g^{(1)}(\mathbf{x}_{12},\tau)|$, is known then the phase difference is all that is required to define relationship between the field points.
But in the case of unknown field distributions, measuring this quantity alone is ambiguous because it cannot distinguish between light states like: entangled photons, incoherent light, or coherent and thermal light (See Table \ref{PhotStates}). The ability to resolve such potential ambiguities
is highly significant, for example, for the current debate over the existence of microcavity-polariton  Bose-Einstein condensation (BEC).\cite{Snoke03}--\cite{Boiko08}

Table \ref{PhotStates}
shows that to distinguish coherent \cite{Glauber63B} and thermal (chaotic) light states \cite{Scarcelli04} or entangled photons and incoherent light state,
the second-order correlation function $g^{(2)}(\mathbf{x}_{12}, \tau) {=} \frac{\langle I_1(t) \, I_2(t+\tau)\rangle}{\langle I_1(t)\rangle \langle I_2(t) \rangle }$,
associated with Hanbury Brown and Twiss (HBT) correlated intensity fluctuations \cite{HBT56}, must also be measured. Here, $I_{1,2}(t)$
is the light intensity at a point $\pm \frac12 \mathbf{x}_{12}$
and time $t$.
However, recalling some of the early controversies surrounding photon-correlation measurements \cite{Adam55,Brannen56,HBT_QE,Parcel56}, reminds us that
second-order correlation measurements are technically more difficult to accomplish in the optical range and requires dedicated single-photon detectors as well as a coincidence-counting apparatus.
An additional difficulty arises due to the impossibility of taking single, simultaneous image of
the $g^{(2)}(x¥,t¥,x,t)$ distribution.
Instead, a pair of detectors and a beam splitter must be used to sample the distribution over a
range of detector-pair positions.
More recent correlation measurements, such as those for BEC cavity polaritons, often require  the ability to resolve the spatial dependence of $g^{(2)}$ in a single run.
Taken together, all these requirements demand an integrated, monolithic, photon detector
(as flexible as a camera), that is capable of imaging intensity noise-correlations.

\begin{table}[fbt]
\caption{\label{PhotStates} Values of first and second order
correlation functions for incoherent, coherent and thermal light
states. Single-mode states are considered, $\theta$ is the angular width of the source, $\lambda$ is the wavelength, $\tau_c$ is the coherence length. Integration effects due to limited detector response times and resolution of coincidence counter are not indicated.}
\begin{center}
\begin{tabular}{lcc}
\hline
Function  & $g^{(1)}(\mathbf x ,\tau )$ & $g^{(2)}(\mathbf x,\tau )$\\
\hline
Incoherent 	& 0	& 1\\
Coherent 	& 1	& 1\\ 
& & \\
Thermal 		& $\text{sinc}\Bigl(\pi \theta\frac x\lambda \Bigr)\,
\exp \Bigl(-\pi \frac{\tau ^2}{2\tau _c^2}\Bigr)$	&
$1+\text{sinc}^2\Bigl(\pi \theta \frac x\lambda \Bigr)\,
\exp \Bigl(-\pi \frac{\tau ^2}{\tau _c^2}\Bigr)$\\
			& 	$[~g^{(1)}(0)=1~]$		& $[~g^{(2)}(0)=2~]$  \\
 & & \\
Entangled	& 0	& $\text{sinc}^2\Bigl(\pi \theta \frac
x\lambda \Bigr)\,\exp \Bigl(-\pi \frac{\tau ^2}{\tau _c^2}\Bigr),$\\
  			& 		& $[~g^{(2)}(0)=1~]$ \\
\hline
\end{tabular}
\end{center}
\end{table}

Elsewhere \cite{Boiko09}, we have presented such a device and applied it in a table-top implementation of a stellar HBT interferometer \cite{HBT56A}, which in their original paper was used to measure the apparent angular diameter of the star Sirius from the current-noise correlations of two detectors separated along a variable  baseline. As seen from the Table \ref{PhotStates}, row for a thermal light source, the angular width of Sirius star $\theta=0.0063''$ assumes that detector separation  baseline $x_{12}$ can be as large as a $10^6 \times \lambda$,  yet enabling  to observe second-order correlations.

However for a table-top implementation, for which
the angular width of a typical source is $\theta \sim 1000''$ (\textit{e.g.}, a 200 $\mu m$ spot seen from a distance of 2 cm),  
the detectors have to be placed at a distance $x_{12}$ no more than a few tens of wavelength $\lambda$ in order to observe correlations. It was thus not possible to measure $g^{(2)}$ in the optical spectral range with stand alone detectors located in one plane.
A conventional table-top implementation
necessitates the introduction of a beam splitter together with two
detectors at equivalent planes related by reflection, as in the original HBT implementation \cite{HBT56}. At the same time for such delicate experiments demanding a clear-cut answer whether the intensity correlations exist or not, as in the polariton BEC experiments, a strong sensitivity to variations in the relative detector position might render the measurement procedure impractical and 
the results biased 
by an error in the relative detector position of the few tens of wavelength scale. Therefore,
we proposed \cite{Boiko09} the use of single photon avalanche detectors (SPADs) integrated into a
monolithic array with a simple data-treatment algorithm for realizing table-top (laboratory) measurements of an HBT interferometer in the original "stellar"-like configuration.

%
%

Here, we report on the peculiarities of our SPAD detector-array for probing
local $g^{(2)}$ correlations. The structure of this paper is as follows.
In Sec.~\ref{sec:imager}, we describe our detector system, which comprises a 4$\times$4
array of Si SPADs implemented in CMOS technology (Fig. \ref{fig1}) and a simple algorithm to treat the multiphoton time-of-arrival distributions from different SPAD pairs, implemented off-chip.
The device also
incorporates on-chip high-bandwidth I/O circuitry which facilitates the external data
processing. We discuss the data treatment procedure to
correct for spurious correlations in measurements (Sec.~\ref{sec:multiphoton}) and
we demonstrate the operation of our device for cases of very
different light statistics (Sec.~\ref{sec:results}):
measuring the beat note of a multimode coherent source,
measuring the depth of intensity oscillations in a superposition of incoherent
and thermal source,
and measuring the instrument response function of a table top interferometer as a function of the source angular-width.

\section{Structure of $g^{(2)}$ imager} \label{sec:imager}

\begin {figure}[tbp]
\centering\includegraphics [width=10cm] {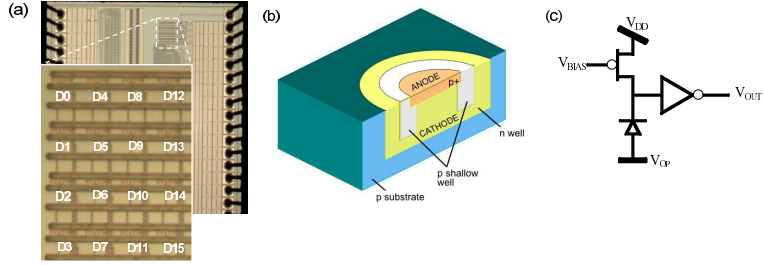} \caption
{Micrograph of the 4$\times$4 SPAD array (a), schematics of the SPAD pixel
structure (b) and electronic readout circuit (c). For a typical operation conditions $V_{\text{OP}}=-21V$, which by 4 V exceeds the breakdown voltage, $V_{\text{DD}}=3.3 V$ and  $V_{\text{BIAS}}=0 V$. } \label{fig1}
\end {figure}

Our $g^{(2)}$ imager is based on a single photon avalanche diode
(SPAD) array \cite{Niclass06}, whose  photomicrograph is shown in
Fig. \ref{fig1} (a). The array pitches are $30$ and $43 \mu m$ in the horizontal and
vertical directions, respectively. The chip is fabricated in a 0.35 $\mu$m
CMOS technology. Each pixel of 4$\times$4
SPAD array is independently accessible
and has a structure 
as
indicated in Fig. \ref{fig1} (b).
The SPADís p$^+$-n-well junction is reverse biased above breakdown by a voltage known as excess bias voltage. In this mode of operation, called Geiger mode, single photons may be detected and counted using auxiliary electronics. The anode is enclosed in the p$^+$-diffusion obtained from the standard implant used to define a PMOS channel. To prevent premature edge breakdown 
in the periphery of the multiplication region, a 3.5 $\mu$m guard ring of lightly doped p-type implant is designed to surround the anode.

The p$^+$ anode of each SPAD pixel is kept at a negative voltage $V_{\text{OP}}$ of 21 V, which by 4 V exceeds the breakdown voltage  (Fig. \ref{fig1} (c)). When an avalanche is triggered, e.g. as a result of photoabsorption in the depleted n-well region or thermally induced carrier release from a trap defect, the avalanche is subsequently quenched
via a resistive current path
in the PMOS transistor structure with the transistor gate kept at a ground potential ( $V_{\text{BIAS}}=0$).
This ballast resistivity
brings the bias voltage across the p$^+$-n junction below the breakdown voltage  and is used to read out the photodetection events.
After avalanche quenching followed by a recharge of the depletion region, the detection cycle is completed and can be initiated again by arriving photons or thermally excited carriers. Each SPAD pixel comprises a high-bandwidth inverter so as the  Geiger pulses are converted into digital signals and transmitted for off-chip data processing.
%
%
This approach considerably improves the signal-to-noise ratio, however as discussed in section \ref{BackgroundCorrSec}, the data should be corrected for spurious correlations due to a possible crosstalk between SPAD pixel outputs. 

During the time required to
complete a detection cycle, the detector is insensitive to
arriving photons. In our structures, the dead time is 12-15 ns, so as  a theoretical limit on the count rate set by detector saturation (pile-up) effects is about 30 MHz.

Tunneling of minority carriers through the shallow p-doped guard ring and/or trap-assisted carrier release may trigger an avalanche and thus
create spurious Geiger pulses. 
The lowest detectable photon flux is set by the dark
count rate (DCR) of SPADs, which at room temperature conditions, is in the 5--10 Hz range. Such low DCR is achieved by implementing n-wells of small diameter $d$=3.5 $\mu$m. The lowest detectable photon flux density in our experiments is thus 10$^{-8}$ photons/s/cm$^2$ and the dynamic response range is 64 dB.

%
%
At 4V excess bias voltage, the measured photon detection probability (PDP) of such 0.35$\mu$m-CMOS SPAD pixels is 40 \% at a wavelength 450 nm and exceeds 25 \% in the wavelength range at 400 - 550 nm. PDP is proportional to excess bias. The spectral sensitivity curve and pixel statistics on dark count rate can be found in \cite{Niclass06}. For our experiments, the chip was mounted on a standard ceramic case and was wire-bounded with 20 $\mu$m-thick aluminum wires of 6-7 mm length spaced by 100 $\mu$m. The bonding wires can be seen on the peripheries of the chip in Fig. Fig. \ref{fig1} (a). For what follows it is important to know that the bonding wires of individual detector data lines are ordered with the SPAD pixel number $i$ indicated on the microphotographic image.





Since all 16 detectors in
the array have separate parallel outputs, $ \bigl( \begin{subarray}{c} 16 \\ 2 \end{subarray} \bigr) = 120$ HBT
interferometer measurements are possible simultaneously between
various detector pairs at temporal resolution set by the SPAD
timing jitter characteristics ($~80ps$). Measurements of pairwise second order correlation are based on implementing expression
\cite{Glauber63A}:
\begin{equation}
g^{(2)}( \mathbf{x}_{ij}, \tau)= \frac{ {\mbox{Tr}} (\hat
\rho \hat a_i^{+} \hat a_j^{+} \hat a_j \hat a_i )}{ {\mbox{Tr}}
(\rho \hat a_i^{+} \hat a_i) {\mbox{Tr}} (\rho \hat a_j^{+} \hat
a_j)}
\label{g2Glauber}
\end{equation}
with a simple algorithm. Here, integers $i,j  \;(i \neq j)$ enumerate detector pixels $D_i$ and $D_j$ in Fig. \ref{fig1},
 $ \hat a_{i,j}$ is the photon annihilation operator at two detectors,  $\hat \rho$ is the density operator for the field and detectors, and trace is taken over the detector and field states \cite{Kelley64}.
At this stage, the pairwise intensity noise correlations $g^{(2)}(x_{ij}, \tau)$  are computed from multiphoton arrivals at different detector pairs of
the array, using a programmable four-channel 6-GHz
bandwidth digital oscilloscope (Wavemaster 8600A, LeCroy) and
implementing a simple expression
\begin{equation}
\tilde{g}^{(2)}_{ij}(\tau)=\tilde{g}^{(2)}(\mathbf{x}_{ij}, \tau)=\frac{NM\sum_{m=0}^M
\sum_{n=-N/2}^{N/2}X_i^{(m)}(n) \wedge
X_j^{(m)}(n + l)}
{\sum_{m=0}^M\sum_{n=-N/2}^{N/2}X_i^{(m)}(n)
\sum_{m'=0}^M
\sum_{n'=-N/2}^{N/2}X_j^{(m')}(n'+l)}
\label{g2Num}
\end{equation}
where
$X_i$ and $X_j$ are discrete random variables whose values $0$
(no event) or $1$ (photon detection) correspond to the binary
data stream emanating from any pair of
detectors $D_i$ and $D_j$, respectively. The spatial lag $x_{ij}$
is set by the separation of the detector pair within the SPAD
array. Time-lag increments $\tau=lT$  are set by multiples of
temporal resolution $T$ (equivalent to the bin width in conventional methods), where $NT$ is the width of the measurement
window and $M$ is the overall number of measurements series.
Note that both addition and the bitwise {\sc and} operator ($\wedge$ in the numerator of
Eq.~\ref{g2Num}) could be implemented easily using on-chip digital electronic circuits.

As can be seen from Eq.~\ref{g2Num}~, the concept of our $g^{(2)}$ imager~\cite{Boiko09} is drastically different to conventional approaches based on measuring start-stop times and computing a timing histogram of delayed photon arrivals at two detectors, followed by renormalization of the histogram according to a statistical hypothesis for the field (e.g., a hypothesis like $g^{(2)} (\tau=0)=1$ or $g^{(2)} (\tau=\pm NT/2)=1$).
Our  $g^{(2)}$-imager algorithm (Eq (\ref{g2Num})) does not suffer from missing detection events in case of subsequent photon arrivals at the same detector and operates correctly with multiphoton arrivals.
Here, we will show that differently from the conventional approach, it operates correctly with
modulated intensity signals at any count rates within the dynamical response range of SPADs and any width $NT$ of the temporal window of interest.

An inherent limitation of our current prototype detector-array is the off-chip data treatment 
implemented
with a general purpose digital oscilloscope that enables user-defined Matlab data processing subroutines.
As a result of the relatively slow Matlab interpreter, a data trace of 5 $\mu$s length (N=5000 points at T=1ns resolution) requires 75 ms of computations to calculate the second-order correlation function in 200 points  (in the interval $-100ns<\tau<100ns$). To achieve a low standard variation of the processed $\tilde g^{(2)}$ data, about M=10$^6$ such data traces needs to be acquired and treated.
The undesirable latency of this numerical processing step could be reducesd dramatically by integrating our robust algorithm into the chip.

\section{Multiphoton correlation measurements} \label{sec:multiphoton}
\label{BackgroundCorrSec}

\begin {figure}[tbp]
\centering\includegraphics[width=12cm] {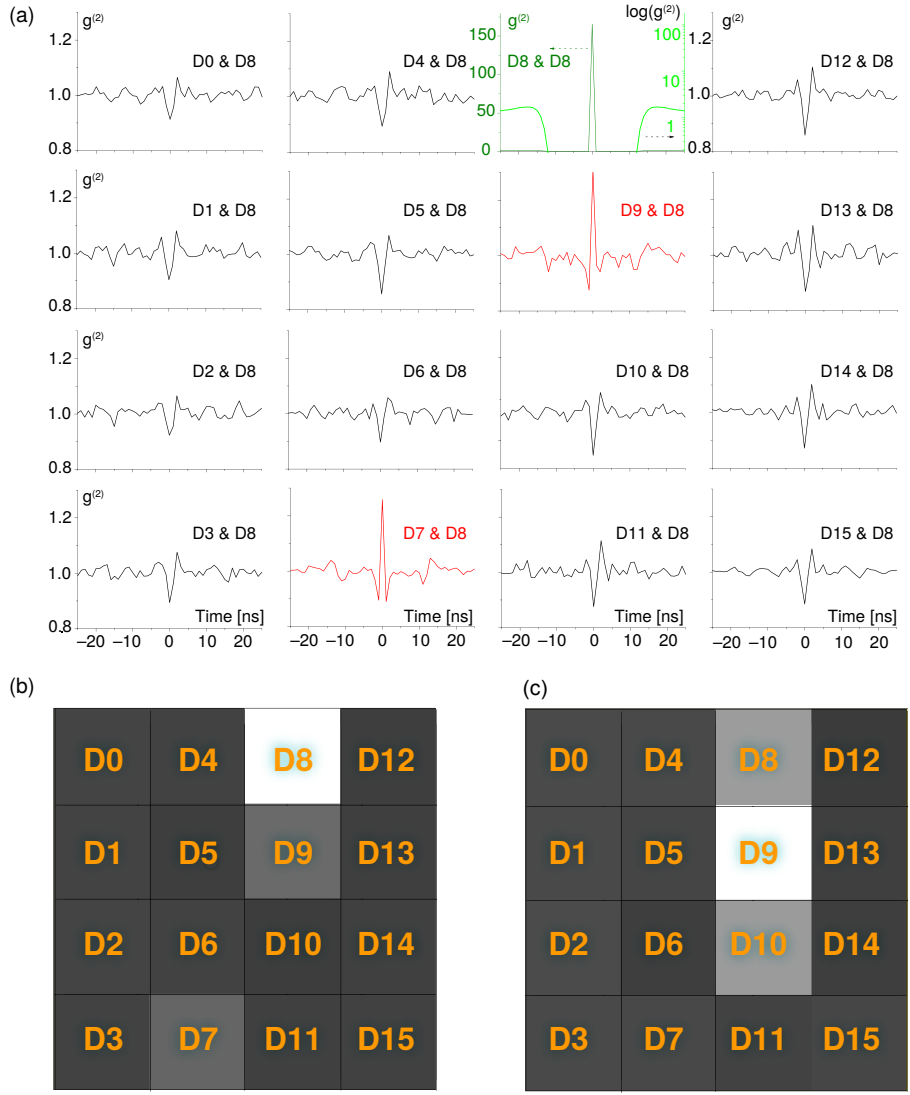} \caption {
Second-order correlation functions $g^{(2)}(\mathbf{x}_{ij}, \tau  ) $ of SPAD array pixels (a) 
and the maps of correlation maxima
$g^{(2)}(\mathbf{x}_{ij}, 0 ) $ [(b) and (c)] measured for incoherent light using
detector D8 [(a) and (b)]  or D9 (c) 
as a reference. The time resolution
is 1 ns. The panels corresponding to individual SPAD pixels are
arranged  in the same order in which they appear in the imager pattern and corresponds to SPAD pixel position in Fig.\ref{fig1}. Green scale curves in (a) 
show autocorrelation curves $g^{(2)}(0 , \tau) $ in linear (olive curve, left axis) and logarithmic (green curve, right axis) scales}\label{Fig2}
\end {figure}

The performance of our $g^{(2)}$ imaging device (Fig. \ref{fig1}) was tested by
measuring the distributions $g^{(2)}(\mathbf{x}_{ij}, \tau )$ when the 4$\times$4 SPAD array detector is uniformly illuminated  with an incandescent light bulb (incoherent broadband light source). Imperfections in $g^{(2)}$ imaging with the SPAD array chip might be caused by several 
reasons including
the concept of monolithic implementation itself. Fig. \ref{Fig2} shows two time-resolved correlation patterns $\tilde g^{(2)}_{ij}(\tau) $ (left panels) and correlation maxima images $\tilde g^{(2)}_{ij}(0) $ (right panels) acquired with respect to detectors D8 (top panels , $j=8$) and D9 (bottom panels,  $j=9$) respectively. In the figure, the pairwise correlations $\tilde g^{(2)}_{i8}(\tau) $ ($\tilde g^{(2)}_{i9}(\tau) $) are ordered according to the detector index $i$ in Fig.\ref{fig1}. For an ideal $g^{(2)}$-imager, one would expect to measure no correlations  ($g^{(2)}(x_{ij}, \tau)=1$) and a uniform correlation maxima map for all pairwise detector combinations. However, Fig. \ref{Fig2} reveals a deviation from the expected distribution.



Measured second-order auto-correlation function of a detector  $\tilde g^{(2)}_{ii}(\tau)$
shows a large coincidence peak $g^{(2)}(0)\gg 1$ at zero time lag [green scale curves in (a)
].  In fact, measurements with one and the same detector at $\tau=0$ do not represent statistical correlations
between field states projected on two independent detectors
in Eq.~\ref{g2Glauber} \cite{Glauber63A,Kelley64}.
Instead, the statistical  algorithm in Eq.~\ref{g2Num} is applied to one and the same field state projection. The excess correlations at zero time-lag can be understood by noting that
the bitwise {\sc and}-ing of the binary data-streams  in Eq.(\ref{g2Num}) 
yields $\langle X_i^{(m)}(n) \wedge
X_i^{(m)}(n)\rangle=\langle X_i^{(m)}(n) \rangle = \mu T$ for the average detection rate $\mu$ and time window $NT$. The result is then $\tilde g^{(2)}_{ii}(0)=\frac{N \dot ( \mu NT)}{(\mu NT)^2 }=1/\mu T$. In Fig \ref{Fig2}, the correlation curves are acquired at resolution time $T$=1 ns and the average rate of detections 6 MHz, yielding an estimate $\tilde g^{(2)}_{jj}(0)=166$,
%
in perfect agreement with the acquired data 
in Fig.\ref{Fig2} (olive curve, left axis).

In the vicinity of the self-correlation peak, the dead-time bands of $\tilde g^{(2)}_{ii}(\tau)=0$ are due to a recharge of the detector depletion region after each Geiger pulse.
Measurements in our SPAD array indicate the dead time $\tau_D$ of 12-15 ns (Fig.\ref{Fig2} (a), 
green curve, right axis). At larger time lags $|\tau|>\tau_D$,  the self-correlation function $\tilde g^{(2)}_{ii}(\tau)$ again shows an excess  of delayed coincidences, exponentially decaying 
to $g^{(2)}=1$. [Note a logarithmic scale of the right axis in (a)
.] This time, i.e., the excess of delayed coincidences, is due to the thermal release of minority carriers from traps in the depletion region, that were
captured during the preceding avalanche current pulse. A casuality relationship between afterpulses and their generating photodetection events explains the measured increase of delayed coincidences for the incoherent light source \cite{Enderlein05,Overbeck98}:
%
\begin{equation}
\tilde{g}^{(2)}_{ii}(\tau )=1 +
\frac{ p_A(|\tau|)}{(1+\varepsilon)^2\mu}, \qquad |\tau|>\tau_D,
\label{g2AfterPulseii}
\end{equation}
where $p_A(t)dt$ is the probability to detect an afterpulse in the time interval $(t,t+dt)$ and $\varepsilon=\int^{\infty}_{\tau_D} p_A(t)dt$ is the overall probability of afterpulsing. We stress that the impact of afterpulsing effects on auto-correlation function (\ref{g2AfterPulseii}) varies with the incident photon flux and is particularly pronounced at small count rates $\mu$.
Measuring auto-correlation curve (\ref{g2AfterPulseii}) for a larger time lag
(not shown in the figures) shows the characteristic decay time of afterpulsing probability 40 ns with an integral effect $\varepsilon=7\%$ at 4 V excess above threshold. The afterpulsing probability $\varepsilon$ is determined by the number of trap defects in the active (n-well) region material, the number of carriers involved in an avalanche and the recharge time. Its impact on (\ref{g2AfterPulseii}) can be drastically reduced by increasing the deadtime of the detector, e.g. by using a smaller recharge current for the depletion region.  We have also noticed a slight decrease of afterpulsing probability with excess of the biasing voltage above the avalanche breakdown. In any case, the auto-correlation measurements, as in (\ref{g2AfterPulseii}),  do not allow the second-order correlation statistics of a source to be accessed but can be used to measure the count rate, dead-time, and afterpulsing probability of SPAD detectors.


The results of pairwise measurements in Fig.\ref{Fig2} deviate slightly
from the expected cross-correlation function 
$g^{(2)}_{ij} (\tau)=1$ as well. For $i\neq j$, the axis scale is limited to the range \mbox{$0.8<g^{(2)}_{ij}<1.3$}. In each of the correlation maps in Figs. (b) and (d), one can distinguish  discrepancies of two kinds: (i) There are a couple of SPAD  pixel-pairs exhibiting a bunching of Geiger pulses at zero time-lag (red curves), while (ii) the rest of detector pairs exhibit anti-correlations at $\tau=0$ (black curves).

Thus in Fig.\ref{Fig2} (a), which was acquired using SPAD pixel D8 as a reference, detectors D7 and D9 exhibit correlation maxima $g^{(2)}(0)\sim 1.3$ measured at temporal resolution $T$=1ns. At higher temporal resolution and shorter integration time ($T$=100ps), its height increases while the natural FWHM measures 320 ps in width (not shown in the figures). Measuring these spurious correlations at various excess voltages above the avalanche threshold and incident light intensities, we have not seen any impact. Such behavior attests for the electrical cross-talk origin of spurious correlations that occur at the SPAD output data lines and have no impact on the detection process.
In particular, the detectors D7 and D9 have data line bonding pads on the CMOS chip adjacent to the wire of SPAD pixel D8. (The bonding wires of
SPAD data
outputs seen at the chip edges in Fig.\ref{fig1} are ordered with the SPAD pixel index $i$.)
These wire lines are subjected to high currents when the logical state changes between 0 (0 V)  and 1 (3.3 V). 
The measured excess of correlations
can thus be attributed to electrical cross-talk between the SPADS pixels with neighboring bonding wires.
This is confirmed in measurements of pairwise correlations with a reference detector D9 (Fig. \ref{Fig2} (c)) showing the excess of correlations for detector pairs D8-D9 and D9-D10. The patterns in Fig.\ref{Fig2} indicate that
there is no direct  optical  crosstalk \cite{Niclass05}  between the
adjacent SPADs within the array, which would
appear in Figs.\ref{Fig2} (b) and (d) as a cross pattern centered at the reference pixel.


All other SPAD pixel pairs in Fig.\ref{Fig2} with larger index difference ($|i-j|\geq 2$) have non-nearest neighbor bonding wires and show small anti-correlations $g^{(2)}_{ij}(0)\sim 0.85$. The anti-bunching dip of Geiger pulses is of 2 ns width followed by strongly damped oscillations of 4 ns period  (they can be seen in Fig.\ref{fig3} (b)). This feature corresponds to reflections in the 90 cm length RF cables transmitting photodetection events data to the processing timing electronics (oscilloscope Wavemaster 8600A, LeCroy). In particular, we attribute this to the impedance mismatch between CMOS output inverters of the SPAD array chip and the 50 Ohm RF cable. Measurements of spurious anti-bunching dips at various illumination power and excess above threshold conditions have not revealed any variations in the dip height or width, attesting the non-detector origin of measured $\tilde g^{(2), bg}_{ij}(\tau)$ bias.

%
%
%
%
%
%
%
%

Spurious correlations and afterpulsing effects have to be taken into account in the photon coincidence measurements utilizing our SPAD arrays. Starting from the results of Ref. \cite{Enderlein05} and spurious background correlations $\tilde g^{(2), bg}_{ij}(\tau)$ between $X_i$ and $X_j$ data lines, one can conclude that
for incoherent and coherent light sources as well as  for thermal, entangled or single photon sources with coherence time smaller
%
%
than the detector dead time $\tau_D$, the measured correlation function (\ref{g2Num}) is
\begin{equation}
\tilde{g}^{(2)}_{ij}(\tau )=1+\frac{g^{(2)}_{ij} (\tau)-1}{(1+\varepsilon)^2}
+(\tilde g^{(2),bg}_{ij}(\tau)-1), \qquad i\neq j
\label{g2AfterPulseij}
\end{equation}
where $g^{(2)}_{ij}$ is the correlation function for  photons that would be measured by ideal detectors. The second term takes a reduction of the photon bunching peak (anti-bunching dip) due to the spurious afterpulses of one detector uncorrelated with photon detection events and afterpulses of another detector. Note that spurious uncorrelated dark counts have the same effect (see also Eq.\ref{g2Superposition}), however for illumination intensity levels used in the experiments discussed here, their impact is negligible.  The third term is the correction due to background correlation caused by data transmission lines from the SPAD chip to the timing electronics.

We shall stress that in agreement with Eq.(\ref{g2AfterPulseij}), there were no  variations observed in the impact of afterpulsing effects on the cross-correlation function with increasing (decreasing) incident light power. 
This feature allows the detector's pairwise correlation measurements (\ref{g2Num}) to be used as an efficient tool for accessing the field statistics $g^{(2)}(\mathbf{x}_{ij}, \tau)$. Unlike conventional detection methods based on start-stop timing
histograms of delayed single photon arrivals, which are not capable to treat multiple photon arrivals at one detector during the start-stop interval measurements, our approach implements properly normalized multiphoton distribution. As such it is robust against missing
detection events, the impact of Poisson-like distribution decay
${\propto} \exp ({-}\mu \tau) $ (see Ref \cite{Boiko09} for details) and intensity
modulation. As opposed to standard methods,  Eq.~\ref{g2Num} permits arbitrary count rates and
temporal window of interest and does not require a statistical
hypothesis to normalize $g^{(2)}$.

\section{Experimental results and discussion} \label{sec:results}


The imager was tested by measuring statistical properties of several light states  including multimode coherent state, near field of an extended intensity-modulated thermal light source and the far field of stationary thermal light source. 

\subsection{Multimode coherent state} \label{sec:multimode}

As a model system for a \textit{ multimode coherent state} we use the emission of a He-Ne laser operating on a
fundamental Gaussian mode of the cavity at 633 nm wavelength transition. 
The cavity length of
$L$=21 cm assumes a large separation of the longitudinal modes  $c/2L$=710 MHz, such that a small optical gain and narrow linewidth 
allows only two or three longitudinal modes to reach the lasing threshold (in function of detuning from the gain line center). No special provision for 
 stabilization of the cavity length or lasing modes has been made.
The relative phases of modes vary rapidly in time such that no intensity modulation at the intermodal frequency can be observed in the emitted output radiation.

The output beam of the laser impinging the  $g^{(2)}$ imaging detector was attenuated to reduce the count rate at detectors down to 2 MHz. The temporal resolution of the timing electronics was set to $T$=100 ps so as to observe dynamics of lasing modes making a cavity roundtrip in 1.4 ns.

\begin {figure}[tbp]
\centering\includegraphics[width=8cm] {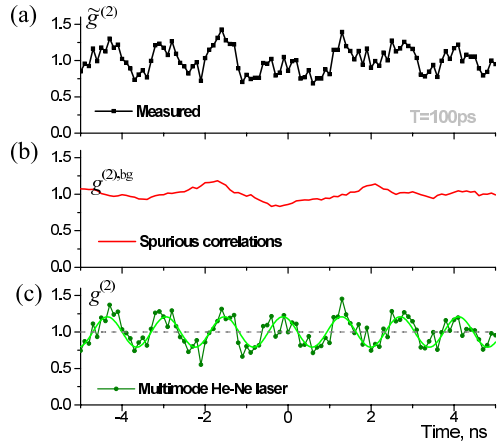} \caption {Multimode
coherent state emitted by 
a He-Ne laser at 632.8 nm wavelength. (a): Acquired second order coherence function $\tilde g^{(2)}_{5,9}(\tau)$ [Eq.(\ref{g2Num})]. (b): Spurious correlations background $g^{(2),bg}_{5,9}(\tau)$ measured with the help of an incandescent light bulb.  (c):
Second order correlation function of the field $g^{(2)}(\mathbf x_{5,9} ,\tau)$   corrected for spurious correlations [Eq.(\ref{g2AfterPulseij})]. The laser cavity length is $L$=21 cm  with the cavity roundtrip time of 1.4 ns (the period of oscillation seen in (c)). The detectors D5 and D9 separated by $x_{5,9}$=30 $\mu$m baseline are used at a temporal resolution of timing electronics $T$=100ps.
}\label {fig3}
\end {figure}

Fig.\ref{fig3} (a) shows the correlation function $\tilde g^{(2)}_{5,9}(\tau)$  [Eq.(\ref{g2Num})] measured by two detectors in the middle of array (detectors D5 and D9) at 30 $\mu$m baseline.  The background component due to a spurious crosstalk at the same detector pair $\tilde g^{(2),bg}_{5,9}(\tau)$, which was measured with an incoherent broadband light source,  is shown in (b). The corresponding correlation function of the multimode laser beam  $g^{(2)}(\mathbf x_{5,9}, \tau)$ calculated from Eq.(\ref{g2AfterPulseij}) after corrections for background correlations and afterpulsing effects in the detectors is plotted in (c). It can be seen that this procedure is an efficient tool for removing spurious (anti-) correlations seen as a dip in both the measured correlations $\tilde g^{(2)}_{ij}(\tau)$ and reference background $\tilde g^{(2),bg}_{ij}(\tau)$.

According to the Table \ref{PhotStates}, 
for each of the modes of the laser, one measures
$g^{(2)}(\tau)=1$. However, as evidenced by the measured second order correlation function in Fig.\ref{fig3} (c),
when these modes are simultaneously excited in the lasing spectrum, one
can see the effective beat signal of intensity and phase noise correlations in the
modes \cite{Grutter69}.
The numerical fit with a cosine function (green curve) reveals the noise beat period of 1.4 ns, that is the time of a roundtrip in the laser cavity. The amplitude of harmonic oscillations 0.2 indicates that the noise of different modes is just partially correlated.    We stress that no such periodic signal is observed in the intensity of the output laser beam.


\subsection{Intensity-modulated extended thermal light source} \label{sec:modulated}

\begin {figure}[tbp]
\centering\includegraphics[width=8cm]{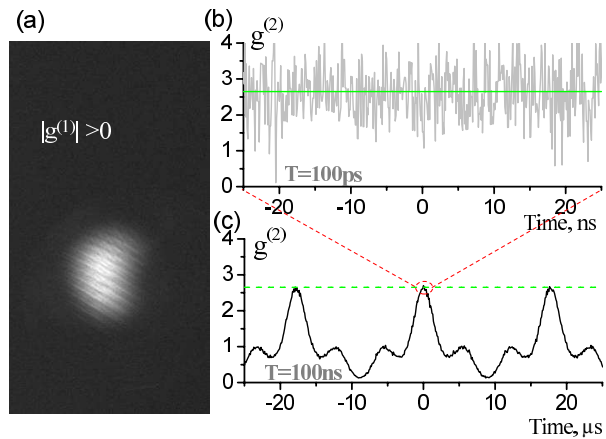} \caption {
Extended quasi-monochromatic thermal light source (546nm line of mercury). (a): Young's interference fringes indicate phase correlation $|g^{(1)}|>0$. (b) and (c):  Measured second order correlation function $g^{(2)}_{5,9} $ 
in the near field of the source obtained after corrections (\ref{g2AfterPulseij})  at temporal resolution 100ps (b) and 100ns (c). The oscillations are due to the AC power supply of the Hg-Ar discharge lamp.
}\label{fig4}
\end {figure}

As a model system for an \textit{extended quasi-monochromatic thermal (chaotic)  light source}, the green line of mercury (546 nm) from the emission spectrum of a low-pressure Hg-Ar discharge lamp (CAL2000, Ocean Optics) is used. The lamp is driven with an original power supply incorporating a 30-KHz AC source producing 1.8 KV pulses to initiate the discharge and 120V in a steady state operation conditions.
The green line of mercury is filtered with a 10 nm band pass filter (FL543.5-
10, Thorlabs).

%
%

Such a quasimonochromatic 
chaotic
light source 
exhibits the first order correlations $|g^{(1)}|>0$ in the Young's interference experiment (Fig.\ref{fig4} (a)). It is 
characterized by
Gaussian distribution of photons in the Glauber
$P$-representation  (Plank distribution in $n$-representation) 
so as one should expect measuring second order correlations $g^{(2)}(x,\tau)=1+\frac 1 q |g^{(1)}(x,\tau)|^2$ with $q$ being the number of photon modes (of equal intensities) reaching the detectors  \cite{Glauber63B}. Here we report on measurements in the near field of the source such that the number of modes is large. (The angular width of the source $\theta \sim$1  in Table \ref{PhotStates}). Therefore, for a detector pair with the baseline of the order of $\lambda$ and more, that is practically in all cases, the second order correlation function should be close to one,  indicating no second order coherence. 

%
%

To measure the second order correlations in the near field of such 
chaotic light source, 
a 3 mm size domain of the lamp discharge region 
was imaged onto the SPAD array at magnification 0.44.
%
%
The image created
by a 150 mm focal distance lens
was overilluminating  
the entire SPAD array
such that we were able to test correlations at 
two pairs of detectors, one in the
middle of the array (SPAD pixels D5 and D9) and one pair with detectors at the opposite array corners (pixels D0 and D15). These detector pairs with baselines of 30 $\mu m$ and 158 $\mu m$ measured correlations $g^{(2)} (x, \tau)$  at the points of the lamp discharge 
that are separated by 68 and 360 $\mu m$ distance,
respectively. 
We were expecting to see no correlations ($g^{(2)}\equiv1$), since the correlation measurements were done 
between \textit{different} emission points of the extended source. However, in both cases, we have observed excess of coincidence counts. Fig.\ref{fig4} details the correlation function $g^{(2)}_{5,9}(\tau)$
measured at the detector pair in the array center. The correlations measured with SPAD pixels at the array corners exhibit the same correlation curve with excess of coincidences $g^{(2)}_{0,15}(0)=2.7$ (not shown in the figure).

The difference between the usual statistical data analysis and our method, can be understood as follows. The distribution $g^{(2)}(\tau)$ is assumed to be an even function about the origin with boundary condition $g^{(2)}(\tau\rightarrow\pm\infty) \sim 1$.
However, when applied to a given
measurement window of interest that is at $\tau=\pm NT/2$, this assumption
can be invalid.
Consider the high resolution data in Fig. \ref{fig4} (b) sampled at $T$=100ps.
Those data, which may be regarded as a uniformly distributed histogram of two-photon arrival times,
can be renormalized to meet
the above boundary condition, viz., $g^{(2)}(\tau)=1$ at the edges of the sample window.
But because of the renormalization, $g^{(2)}$ will not have the required value of $2.7$.


%
%
%

By limiting the temporal resolution to $T$=100 ns and increasing the width of  the measurement window to $NT$=50 $\mu$s, the origin of the excessive correlations can be made visible [Fig. \ref{fig4} (c)]. The measured second order correlation function just reveals the fact of long-period intensity oscillations of the lamp at the double frequency of the AC power supply, at about 60 KHz.
Because of these oscillations and the large angular width of the
source, the photon bunching at $\tau=0$ due to
Planck's statics \cite{HBT56} is not visible.
The change in the width of the measurement window and the temporal resolution (in Figs. (b) and (c) ) has no impact on the measured excess value  $g^{(2)}(0)=2.7$, attesting for correct normalization of the correlation function when multi-photon arrivals are treated with the algorithm (\ref{g2Num})-(\ref{g2AfterPulseij}). Interestingly,
after correction for the afterpulsing effects at the detectors [$\epsilon$ in  Eq.(\ref{g2AfterPulseij})], the minima of $g^{(2)}(\tau)$ function are close to zero, in agreement with fluorescence lifetime
considerations.

%
%
%

\begin {figure}[tbp]
\centering\includegraphics[width=10cm]{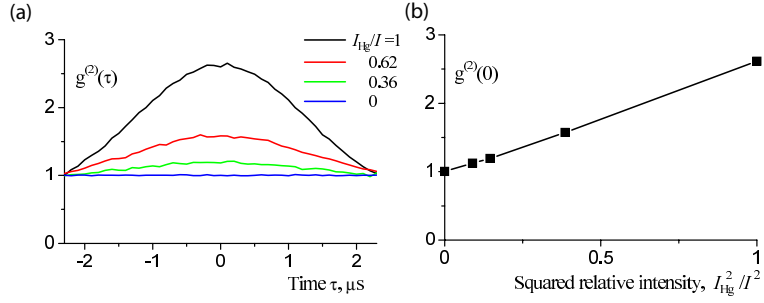} \caption {
Intensity correlations measured for the superposition of incoherent light (from  incandescent lamp) and intensity-modulated quasimonochromatic thermal light (green 546nm line of mercury from Hg-Ar lamp CAL2000) at various relative intensities. Temporal resolution is $T$=100ns. (a): Second order correlation function $g^{(2)}(\tau)$. (b): Correlation function maximum $g^{(2)}(0)$ plotted as squared relative intensity of thermal light incident on detector pair.
}\label {fig5}
\end {figure}

Modulation depth of the correlation function can be tailored in a mixed state. Fig. \ref{fig5}  details second-order correlation measurements for a superposition of incoherent light (as in Fig.\ref{Fig2}) and a quasimonochromatic chaotic light beam (as in Fig.\ref{fig4}) 
at various probabilities to detect photons in 
chaotic light state. 
The incandescent lamp is used as a source of incoherent emission, 
 which is superimposed on mercury green line fluorescence from the intensity-modulated Hg-Ar lamp. The probability $P(Hg)$ that a detected photon has been emitted by mercury atoms, is given by the intensity ratio $I_{Hg}/I$ estimated from the count rates at the detectors. For example,
$P(Hg) = 0.38$ when $I_{Hg}=6.8$ kHz, $I=17.8 kHz$.


%

One can easily show that for a superposition of mutually uncorrelated
photonic
states, characterized by partial intensities $I^{(\alpha)}$ and  correlations $g^{(2),\alpha}(\mathbf{x}_{ij},\tau)$, the second order coherence function   $g^{(2)}(\mathbf{x}_{ij}, \tau) {=} \frac{\langle I_i(t) \, I_j(
t+\tau)\rangle}{\langle I_i(t)\rangle \langle I_j(t) \rangle }$ for the overall photon fluxes $I_{i,j}=\sum_\alpha I^{(\alpha)}_{i,j}$ impinging the detectors assumes the expansion :
\begin{equation}
g^{(2)}_{ij}(\tau )=1+\sum_\alpha (g^{(2),\alpha}_{ij} (\tau)-1)\frac{\langle I^{(\alpha)}_i\rangle \langle I^{(\alpha)}_j\rangle}{\langle I_i\rangle \langle I_j\rangle}
\label{g2Superposition}
\end{equation}
%
%
%
This expression can be simplified  in the case of equal count rates of the detectors, yielding the partial weights $\langle I^{(\alpha)} \rangle ^2/ \langle I \rangle^2$ for state contributions of $g^{(2),\alpha}(\tau)-1$ to the overall excess (or lack) of coincidences of the correlation function. For example, for the overall rate $\langle I \rangle =(1+\epsilon) \langle I^{(\alpha)} \rangle $ of photodetection events caused by a process of intensity $\langle I^{(\alpha)} \rangle $ and statistics $g^{(2),\alpha}_{ij}(\tau)$ accompanied by spurious (incoherent) afterpulses occurring 
with probability $\epsilon$, Eq.(\ref{g2Superposition}) predicts a reduction of the bunching peak (or antibunching dip). The second order coherence function of the entire process is then $1+(g^{(2),\alpha}_{ij}(\tau)-1)/(1+\epsilon)^2$, in agreement with Eq.\ref{g2AfterPulseij}. Along the same lines, the effect of spurious dark counts of detectors can be taken into account at low count rates $\langle I_{i,j} \rangle$.

%

For the experimental results reported in Fig.\ref{fig5}, the incoherent light constituent does not contribute to the excess of correlation (the correlation function $g^{(2),incoh}(\tau)\equiv1$). The  average intensities 
at the detectors (SPAD pixels D5 and D9) are equal so as the contribution from the  chaotic light source of Fig.\ref{fig4} is reduced as $1+(g^{(2),Hg}_{5,9}(\tau)-1) \langle I_{Hg} \rangle^2/\langle I \rangle^2$ [Fig.\ref{fig5} (a)].
The maximum values of the second order correlations $g^{(2)}_{5,9}(0)-1$
linearly grows with the squared
probability of detecting a thermal light photons $\langle I_{Hg} \rangle ^2/\langle I \rangle ^2$ [Fig.\ref{fig5} (b)].

%
%
%

\subsection{Quasi-monochromatic thermal light source}  \label{sec:monochromatic}

\begin {figure}[tbp]
\centering\includegraphics[width=13cm]{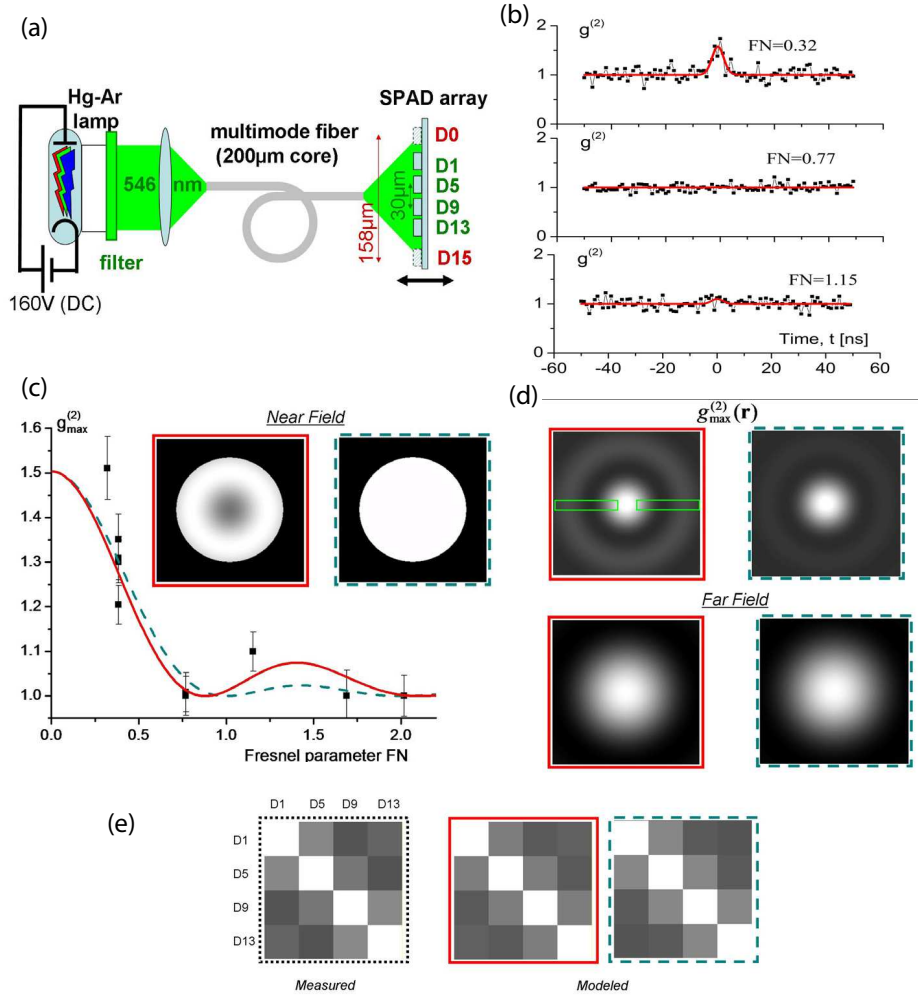} \caption {(a)
Experimental setup of the table-top stellar HBT interferometer.
(b) Correlations measured at various detector separation $x_{ij}$ and distance $L$ to the fiber end. The Fresnel parameter $FN=\frac{w x_{ij}}{\lambda L}$ is indicated in the panels. (c) Measured (points)
and modeled (curves) second-order correlations in function of detector separation for the model Eq.\ref{g1full} (dashed blue curve) and Eq.~\ref{g2fullCorr} (solid red curve).
The inset shows the corresponding near field distributions at the fiber end (indicated with same type ). (d) shows the corresponding modelled $g^{(2)_max}(\mathbf r)$ patterns (top line of panels) and far field patterns (bottom line of panels). (e) Imaged
second-order correlation maxima along the row of the array. Its position in the $g^{(2)}$ plane is indicated in (d) with green solid lines. It is assumed that
$g^{(2)} = 2$ along the diagonal.
The green line of Hg (546nm) is used. Temporal resolution $T=1$ ns.
}\label{fig6}
\end {figure}

So far we have not yet observed the HBT photon bunching in the emission from a \textit{quasi-monochromatic chaotic light source}. To make it 
visible, we replace the power source of the lamp and conduct measurements in the far field. In particular, 
we drive the  Hg-Ar lamp bulb (the same as used in previous experiments) from a DC voltage source.
This lamp bulb (CAL-2000-B, Ocean Optics) having a cold cathode and U-folded
discharge and designed for operation in the AC regime, also operated well with a DC power supply (160V @ 15 mA). To start the discharge, we used the original AC power supply of the lamp, 
which was connected in parallel with a DC source via a filter (2H inductance) such that the AC supply was gradually turned off, whereas the DC source was gradually turned on.

As in the experiment of Fig.\ref{fig4}, the light emitted by the lamp was transmitted through a
10nm-width bandpass filter, which keeps only the emission at the
green line of mercury (546 nm). The inherent difficulty is related to the fact that 
the number $q$ of uncorrelated photon modes contributing to the intensity interference at SPAD array detectors $g^{(2)}(x,\tau)=1+\frac 1 q |g^{(1)}(x,\tau)|^2$ has to be kept small. (The requirement $q \sim 1$ can be also seen from Eq.(\ref{g2Superposition}) for a superposition of $q$ modes of equal intensities.)  For a light beam  of transversal size $w$ seen from detectors at a distance $L$ in the ray cone of angle $\Theta=x_{12}/L$, the number of
photon modes is $q=\int\frac {d^2\mathbf r d^2 \mathbf k}{(2 \pi)^2}\sim\frac1{64} ( k w \Theta )^2=\frac 1 {16} (\pi \frac{w x_{12}}{\lambda L})^2$. For a SPAD array of a few $ 100 \times 100$ $ \mu$m$^2$ size, which defines a characteristic baseline $x_{12}$ of detectors, the requirement $q \sim 1$ imposes the angular width of such extended source to be $\theta=w/L \sim 10^{-2}$ or less (see also Table \ref{PhotStates}).

To compromise the brightness of the source and its angular width at the detectors, the light from the lamp bulb was injected into
a multimode fiber of 1 m length and core diameter $w\sim 200$ $\mu$m [Fig.\ref{fig6}(a) ].  The other end
of the fiber was used to illuminate the SPAD array in the
far field zone of the fiber end, at a variable distance $L$ (1 - 3 cm) from
the fiber such that the whole 4$\times$4 SPAD array is
over illuminated.
Such extended thermal light source is of the small
angular width $w/L{=}10^{-2}$ rad
and exhibits first-order
correlations $g^{(1)}$, when the SPAD array is
replaced by Young double-pinhole interferometer as in Fig. \ref{fig4} (a), which was taken in the same setup configuration but with the Hg-Ar lamp driven from the AC power supply.

In intensity interference measurements with a quasi monochromatic thermal
light source, the coherence width and Doppler-broadened
spectral line have the same impact as in
the case of the Young double slit interferometer.
%
%
The classical expression for second-order spatio-temporal correlations for a non-polarized
single-mode chaotic 
light source are determined by the first-order
correlations~\cite{Glauber63B,Glauber63A,Glauber07}  with coherence
time $\tau_c{=}2\sqrt{2 \pi \ln2}/\Delta \omega$ due to the
inhomogeneous broadening $\Delta \omega$ (FWHM) of the emission
line
\begin{equation}
g^{(2)}(\mathbf x_{ij}, \tau) =1+\frac 12 \left|g^{(1)}(\mathbf
x_{ij},\tau)\right|^2  \\
=1+\frac 12{\rm sinc}^2 \Bigl(\frac{\pi w}{\lambda L} x_{ij}
\Bigr) \, \exp \Bigl(-\pi \frac{\tau^2}{\tau_c^2} \Bigr) \, ,
\label{g1full}
\end{equation}
where the second term in the right hand side takes into account
the decorrelation effects due to unpolarized light (the
coefficient $1/2$), the zero-delay degree of spatial coherence and
the Gaussian profile of the delayed first-order correlation
function for an inhomogeneously-broadened line.

Fig.\ref{fig6} (b) shows several traces $g^{(2)}(\tau)$ measured at different distances to the fiber end at several detector pairs (the Fresnel like parameter $FN= \frac{w x_{ij}}{\lambda L}$ is indicated for each trace). At small $FN<0.5$ (large distance to detectors and small baseline), the detector counts show a correlation peak. At $FN=1$, it disappears and then at $FN\sim1.2$, it   reappears but with a smaller amplitude of correlation excess. At first sight, such behavior corresponds well to the expected oscillations of the sinc function in Eq.(\ref{g1full}). However
measuring correlations at various distances from the fiber end with several detector pairs and correcting the data for afterpulsing effect (Fig.\ref{fig6} (c), points), we find large discrepancies with expected dependence (Fig.\ref{fig6} (c), dashed blue curve). The difference is outside the instrumental error and the fast growth of  $g^{(2)}_{max}$ at $FN\sim1.2$ cannot be compensated for with a numerical fit to the model in Eq.(\ref{g1full}).


To interpret the experimental results in Fig.\ref{fig6}, one shall admit that $g(2)$, being a function of two variables, can be tailored not only in the time domain, as exemplified in previous sections, but it can also be tailored via spatial modulation of the near field distribution. This fact is often  omitted in experiments on photon bunching, leading to wrong data interpretation. In the same way as the time domain variations of intensity (intensity noise) renders invalid the exponential term of the famous expression (\ref{g1full}), the sinc term in that expression is obsolete if the source has a nonuniform intensity distribution in the near field.

Using Glauber' expression $g^{(2)}_{12}=1+G^{(1)}_{12} G^{(1)}_{21}/G^{(1)}_{11}G^{(1)}_{22} $ \cite{Glauber63B} with $G^{(1)}_{ij}= \langle E_{i}E^{*}_{j} \rangle$ for an extended source with near field intensity distribution $I(x,y)$ and detector pair with baseline $x_{ij}$ 
centered at $x=0$, after integration over the wavelet contributions $E_{1,2}\propto \frac {1}{\lambda L} |E_{0}(x,y)|^2  {\rm exp}(-ikx_{12}x/L) dxdy $ from mutually incoherent elements $dxdy$  of the extended source, one finds
\begin{equation}
g^{(2)}(\mathbf x_{ij}, \tau) =1+ \frac 12 \Biggl|\frac {\int I_0(x,y) {\rm exp}   \bigl(-i\frac{2\pi x }{\lambda L} x_{ij}
\bigr)dxdy}{\int I_0(x,y)dxdy}\Biggr|^2
%
%
\exp \Bigl(-\pi \frac{\tau^2}{\tau_c^2} \Bigr) \, ,
\label{g2full}
\end{equation}
which is nothing else than the Van Cittert-Zernike theorem applied for visibility of the second order correlations. Here, as in (\ref{g1full}), the coefficient $1/2$ takes the decorrelation effects due to unpolarized light into account.  We shall stress that only the 1-D Fourier transform of the near field intensity in the direction of the detector baseline is involved. For a uniform intensity distribution, integration over the source area from $-w/2$ to $w/2$ yields expression (\ref{g1full}).

Field distribution in multimode fiber highly depends on excitation conditions and uniformity of the fiber core material \cite{Zhang01,Oh07,Chaikina05}. Due to cylindrical symmetry of the fiber, the intensity distribution has an intensity bump or a dip at the core center. To simplify analysis, we consider only the main radial harmonic of frequency $\Omega$, assuming the approximation $I(r)\propto 1+\gamma cos(\Omega r)$:
\begin{equation}
g^{(2)}(\mathbf x_{ij}, \tau) =1+\frac 12 \Biggl|
\frac{{\rm sinc}\Bigl(\pi FN \Bigr)+\frac {\gamma}2 {\rm sinc}\Bigl( \pi FN
+\frac 12 \Omega w \Bigr)+\frac {\gamma}2 {\rm sinc}\Bigl(\pi FN
-\frac 12 \Omega w \Bigr)}
{1+\gamma {\rm sinc}(\frac 12 \Omega w ) }
\Biggl|^2  \exp \Bigl(-\pi \frac{\tau^2}{\tau_c^2} \Bigr)
\label{g2fullCorr}
\end{equation}
where $\gamma$ is the intensity modulation depth and we have used Fresnel parameter $FN=w x_{ij}/\lambda L$ to shorten the expression. The solid red curve in Fig.\ref{fig6} (c) shows the result of the numerical fit with the model  (\ref{g2fullCorr}), reporting $\gamma=-0.3$ and spatial frequency $\Omega=2.8~\pi/w$. The corresponding near field (shown in the inset) has a small intensity dip at the center. Note that the difference to uniform intensity distribution is remarkable only in the correlation maxima distribution $g^{(2)}_{max}(\mathbf r)$ (Fig.\ref{fig6} (d), top line panels) while it is not visible at all in the far field distribution impinging the detectors (Fig.\ref{fig6} (d), bottom line panels).

Being limited by the number of acquisition channels, we were able to record simultaneous
correlations between four independent detectors form one row of the SPAD array. In Fig.\ref{fig6} (e), $g^{(2)}(x_{ij}, 0)$ measured along the
array row is plotted as a pairwise correlation map $g^{(2)}(i, j )$ for measured (left panel) and calculated with the models (\ref{g2fullCorr}) and (\ref{g1full}) correlation maxima (two right panels). In this image map, the spatial
oscillations of the coherence factor are clearly visible. The measured excess of correlations for the corner pixels on the secondary diagonal corresponds better to the model  (\ref{g2fullCorr}), confirming previous conclusion on impact of the spatial modulation of the light source.

\section{Conclusion}

We have presented a $g^(2)$-imager built with conventional CMOS technology,
which is capable of measuring second-order spatio-temporal correlated photons.
We have discussed  several important aspects related to temporal and spatial modulation of the source intensity when conducting HBT correlation measurements with such imager. Such detectors will find various applications. This approach allows the functions $g^{(2n)}$ of other even
orders to be implemented as well.


The application in Sec. \ref{sec:monochromatic} used a green line emission of mercury in a Hg-Ar discharge lamp as a reliable
source of HBT photons pairs. The lamp was emitting around hundred $\mu$W of thermal light at the green-line transition, 
only small fraction of which can be coupled into a multimode fiber. Yet the photon flux $\mu$ emanating from the fiber was quite high, about 10$^{11}$ photons/sec. If not decorrelations due to the angular width of such source,  one would need detectors and coincidence electronics with integration time $T\sim$10ps to detect coincidences at the highest rate $R_c \sim 2 \mu ^2 T$ of about 10$^{11}$ photon pairs/sec in the near field of the fiber.

%
%
In practice, because of the low collection efficiency of detectors in the far field of the fiber end, the count 
and coincidence rates 
are very low, about $\mu$=100  kHz and $R_c$=10 Hz respectively, for integration time $T$=1 ns. Using detectors of large 
diameter $d$ 
will not overcome the problem, yielding a reduction of measured excess of correlations $(g^{(2)}(\tau)-1)$ by a factor of
$\sim w^2d^2/\lambda^2 L^2 $ [see Sec.\ref{sec:monochromatic}  ],
like long integration time $T$ results in a decoherence factor of $\tau_c^2/(\tau_c^2+T^2)\sim \tau_c^2/T^2$ \cite{Scarl66}.

Although the alternative of building a quasi-thermal source by using a He-Ne laser beam impinging a rotating sintered disk offers higher intensity, larger coherence time, and hence much faster acquisition, we scrupulously avoided this type of source so as not to introduce the undesirable
possibility of spurious correlations such that to observe a clear impact of detector imperfections, non-stationarity and non-uniformity of the field on measured second-order correlations.

It is interesting to stress 
the impossibility of detecting a normalized correlation function $g^{(2)}(\tau)$ for entangled biphotons (Table \ref{PhotStates}) \cite{Scarcelli04}. A state of the art entangled photon source based on spontaneous parametric down conversion in periodically-poled nonlinear crystal shows conversion efficiency of 10$^{-7}$ and is capable of producing bi-photons at rates of 10$^8$ photon pairs/sec per miliwatt of pump power.  The principal difference to a thermal light source illuminating detectors is that all generated bi-photon pairs produce coincidence detections, provided  perfect collection efficiency and
no decoherence due to the angular width of the source.
%
%
%
Technical difficulties arise from low overall power 
and a very short coherence time of bi-photons (typically, $\tau_c \sim$0.1 ps) so as 
to measure the peak value $g^{(2)}(0)=1$ for entangled photons (see Table \ref{PhotStates}) 
one will need detectors with the response time $T\sim$0.1 ps, which do not exist yet. On practice, one just measures non-normalized second order correlations $G^{(2)}(\tau)$  with the time lag $\tau$ defined 
by the optical path difference to two detectors.

Future work will include the development of larger arrays of SPADs, the integration of on-chip
data processing based on equation (1), and the extension to other $g^{(2)}$-imaging applications.

%
%

\section*{Acknowledgements}
We would like to acknowledge Christian Depeursinge fabricated the double pinhole for interferometeric measurements.
This research was supported, in part, by a grant of the Swiss
National Science Foundation.

%
%
%
%


\begin{thebibliography}{99}

\bibitem{Lundeberg07} L. D. A. Lundeberg, G. P. Lousberg, D.L. Boiko, E. Kapon, ``Spatial coherence measurements in arrays of coupled vertical cavity surface emitting lasers,'' Appl. Phys. Lett. {\bf 90}, 021103--3 (2007)

\bibitem{Snoke03}D.~W.~Snoke,
``When should we say we have observed Bose condensation of
excitons?'' Phys. Stat. Sol. (b) \textbf{238}, 389--396 (2003).


\bibitem{Deng02} H.~Deng, G.~Weihs, C.~Santori, J.~Bloch,
Y.~Yamamoto, ``Condensation of Semiconductor Microcavity Exciton
Polaritons,'' Science \textbf{298}, 199--202 (2002)

\bibitem{Deng06}H. Deng,
D. Press, S. G\"otzinger, G. S. Solomon, R. Hey, K. H. Ploog, Y. Yamamoto,
``Quantum Degenerate Exciton-Polaritons in Thermal Equilibrium,''
Phys. Rev. Lett. \textbf{97}, 146402--4, (2006).


\bibitem{Kasprzak06}J.~Kasprzak, M. Richard, S. Kundermann, A. Baas, P. Jeambrun, J. M. J. Keeling, F. M. Marchetti, M. H. Szyma\'nska, R. Andr\'e, J. L. Staehli, V. Savona, P. B. Littlewood, B. Deveaud and Le Si Dang,
``Bose-Einstein condensation of exciton polaritons,'' Nature
\textbf{443}, 409--414 (2006).

\bibitem{Christopoulos07}S.~Christopoulos,
G. Baldassarri H\"oger von H\"ogersthal, A. J. D. Grundy,
P. G. Lagoudakis, A.V. Kavokin, and J. J. Baumberg,
G. Christmann, R. Butt\'e, E. Feltin, J.-F. Carlin, and N. Grandjean,
``Room-Temperature Polariton Lasing in Semiconductor
Microcavities,'' Phys. Rev. Lett.  \textbf{98}, 126405--4, (2007).

\bibitem{Bajoni07}
D. Bajoni, P. Senellart, A. Lema\^itre, and J. Bloch, ``Photon lasing in GaAs microcavity: Similarities with a polariton condensate,'' Phys. Rev. B \textbf{76}, 201305(R)--4 (2007)


\bibitem{Bloch08}J.~Bloch, D. Bajoni, P. Senellart, E. Wertz, I. Sagnes, A. Miard, and A. Lema\^itre,``Polariton quantum degeneracy in GaAs microcavities,'' presented at the
2008 Latsis Symposium at EPFL on Bose Einstein Condensation in
dilute atomic gases and in condensed matter,
Lausanne, Switzerland,28-30 Januarry 2008. 

\bibitem{Balili07} R. Balili, V. Hartwell, D. Snoke, L. Pfeiffer, and  K.
West, ``Bose-Einstein Condensation of Microcavity Polaritons in a
Trap,'' Science \textbf{316}, 1007--1010 (2007).

\bibitem{Boiko08} D.L. Boiko, ``Towards r-space Bose-Einstein condensation of photonic crystal exciton polaritons,'' in {\it Proceedings of the Progress in Electromagnetics Research Symposium PIERS 2008},  (Cambridge MA, USA, July 2-6, 2008), pp 659-665 (2008); {\it idem}, PIERS Online \textbf{4}, 831-837 (2008).  

\bibitem{Glauber63B}R.~J.~Glauber,
``Coherent and incoherent states of the radiation field,'' Phys.
Rev. \textbf{131}, 2766--2788, (1963).

\bibitem{Scarcelli04} G. Scarcelli, A. Valencia and Y. Shih, ``Two-photon interference with thermal light,''  Europhys. Lett., \textbf{68} , pp. 618--624 (2004).

\bibitem{HBT56}R.~Hanbury~Brown and R.~Q.~Twiss,
``Correlation between photons in two coherent beams of light,''
Nature \textbf{177}, 27--29 (1956).

\bibitem{Adam55} A. Ad\'am, L. J\'anossy, R. Varga, ``Coincidences between photons contained in coherent light rays,'' Acta Physica Hungarica \textbf{4}, 301--315 (1955)

\bibitem{Brannen56} E. Brannen and H. I. S. Ferguson, ''The question of correlation between photons in coherent light beams,'' Nature \textbf{178}, 481--482 (1956)

\bibitem{HBT_QE} R.~Hanbury~Brown and R.~Q.~Twiss, ``The question of corelation between photons in coherent light rays,'' Nature \textbf{178}, 1447-1448 (1956),

\bibitem{Parcel56} E. M. Purcell, ``The question of corelation between photons in coherent light rays,'' Nature \textbf{178}, 1449-1450 (1956)



\bibitem{Boiko09} D. L. Boiko, N. J. Gunther, N. Brauer, M. Sergio, C. Niclass, G. B. Beretta and E. Charbon, ``A quantum imager for intensity correlated photons,''   New J. Phys. \textbf{11} 013001--7  
    (2009)




\bibitem{HBT56A}R.~Hanbury~Brown and R.~Q.~Twiss,
``A test of a new type of stellar interferometer on Sirius,''
Nature \textbf{178}, 1046--1048 (1956).


\bibitem{Niclass06}C. Niclass, M. Sergio and E Charbon, ``A Single Photon Avalanche Diode Array Fabricated in 0.35 $\mu$m CMOS and based on an Event-Driven Readout for TCSPC Experiments'' in {\it Proc. SPIE Opt. East (Boston)} vol \textbf{6372} (Bellingham, WA, SPIE
Optical Engineering Press, 2006) p 63720S-12.




\bibitem{Glauber63A}R.~J.~Glauber,
``The Quantum Theory of Optical Coherence,'' Phys. Rev.
\textbf{130}, 2529--2539 (1963).


\bibitem{Kelley64} P. L. Kelley, W. H. Kleiner, ``Theory of Electromagnetic Field
Measurement and Photoelectron Counting,'' Phys. Rev.  \textbf{136}
A316--A334 (1964).



\bibitem{Enderlein05}Enderlein J.; Gregor I, ``Using fluorescence lifetime for
discriminating detector afterpulsing in fluorescence-correlation
spectroscopy,'' Rev. Sci. Instrum. \textbf{76}, 033102--5 (2005).

\bibitem{Overbeck98} E. Overbeck and C. Sinn, ``Silicon avalanche photodiodes as
detectors for photon correlation experiments,'' Rev. Sci. Instrum.
\textbf{69}, 3515--3523 (1998).

\bibitem{Niclass05}
C. Niclass, A. Rochas, P.A. Besse, and E. Charbon, ``Design and
Characterization of a CMOS 3-D Image Sensor Based on Single
Photon Avalanche Diodes,'' IEEE J. Solid-State Circuits
\textbf{40}, 1847--1854 (2005).


\bibitem{Glauber07}R.~J.~Glauber,
``Nobel Lecture: One hundred years of light quanta,'' Ann. Phys.
(Leipzig) \textbf{16},
6--24 (2007),

\bibitem{Zhang01} J.Zhang, Q. Li, W. Pan, Y. Chen, ''Ring-Shaped Field Pattern: The Fundamental
Mode of a Multimode Optical Fiber,'' Fiber and Integrated Optics, \textbf{20}, 403-410 (2001).

\bibitem{Oh07} C. W. Oh, S. Moon, Suhas P. Veetil, D. Y. Kim, ``An angular offset launching technique for bandwidth enhancement in multimode fiber links,'' Microwave and Optical Technology Lett. \textbf{50}, 165--168 (2007).

\bibitem{Chaikina05} E. I. Chaikina, S. Stepanov, and A. G. Navarrete, E. R. M\'endez, T. A. Leskova, ``Formation of angular power profile via ballistic light transport in multimode optical fibers with corrugated surfaces,'' Phys. Rev. B \textbf{71}, 085419--9 (2005).

\bibitem{Grutter69}
A. A. Gr\"utter, H. P. Weber, and R. D\"andliker, ``Imperfectly Mode-Locked Laser Emission and Its
Effects on Nonlinear Optics,'' Phys. Rev. \textbf{185}, 629--643 (1969).

\bibitem{Scarl66}D. B. Scarl, ``Measurement Of Photon Time-Of-Arrival Distribution In Partially
Coherent Light,'' Phys. Rev. Lett.\textbf{17} 663--666 (1966).


\end{thebibliography}
\end{document}